\documentclass[12pt]{article}
\usepackage{color}
\usepackage{amsmath,amssymb}
\usepackage{amscd}
\usepackage[pdftex]{graphicx} 
\usepackage{latexsym}
\usepackage{cite}

\allowdisplaybreaks[1]

\newcommand{\vev}[1]{\left\langle #1 \right\rangle}

\newcommand{\del}{\partial}

\DeclareMathOperator{\sgn}{sgn}

\newcommand{\blueflag}[1]{{\color{blue} #1}}

\renewcommand{\thefootnote}{\fnsymbol{footnote}}
\makeatletter
\@addtoreset{equation}{section}

\begin{document}


\begin{titlepage}
\thispagestyle{empty} 
\begin{flushright}
\end{flushright}

%
%

\begin{center}
\noindent{\large \textbf{
   Hybrid Quantum Annealing via Molecular Dynamics
 }}
\end{center}

\vspace{0.2cm}

\begin{center}
\noindent{Hirotaka Irie,$^{*,a,b}$ 
Haozhao Liang,$^{c,d}$ 
Takumi Doi,$^{b,c}$ 
Shinya Gongyo,$^{b,c}$ \\
and Tetsuo Hatsuda$^{b}$
}
\end{center}

\vspace{.0cm}

\begin{center}
{\em
AI R\&I Division, Advanced Research and Innovation Center, \\
DENSO CORPORATION, Global R \& D Tokyo, Tokyo 108-0075, Japan$^{a}$ 
 \\
}
\vspace{0.15cm}
{\em
RIKEN Interdisciplinary Theoretical and mathematical Sciences Program \\
(iTHEMS), Saitama 351-0198, Japan$^{b}$
 \\
}
\vspace{0.15cm}
{\em
RIKEN Nishina Center (RNC), Saitama 351-0198, Japan$^{c}$
\\
}
\vspace{0.15cm}
{\em
Department of Physics, Graduate School of Science, \\
The University of Tokyo, Tokyo 113-0033, Japan$^{d}$ \\
} 
\vspace{0.15cm}
\footnotesize ${}^*$Correspondence to {\tt hirotaka.irie.j3a@jp.denso.com}
\end{center}

\vspace{0.2cm}

\begin{abstract}
A novel quantum-classical hybrid scheme is proposed to efficiently solve large-scale combinatorial optimization problems.
The key concept is to introduce a Hamiltonian dynamics  of the  
 classical flux variables associated with the quantum spins of the transverse-field Ising model.
  Molecular dynamics  of the classical fluxes can be used as a 
powerful preconditioner to sort out the frozen and ambivalent spins for quantum annealers.
 The performance and accuracy  of our smooth hybridization in comparison to the standard
  classical algorithms (the tabu search and the simulated annealing) 
  are demonstrated by employing the MAX-CUT and Ising spin-glass problems.
\end{abstract}

\end{titlepage}

\newpage

\renewcommand{\thefootnote}{\arabic{footnote}}
\setcounter{footnote}{0}

\noindent
{\bf {\large Introduction}}\\

\noindent

Combinatorial optimizations are ubiquitous and generally represented by the Ising spin-glass model, which is computationally classified as an NP-hard problem \cite{SpinGlassIsingNPhard}.
The quantum annealing with  a transverse-field Ising model \cite{Kadowaki:1998,Morita:2008} as well as the adiabatic quantum computation \cite{Farhi:2001,Aharonov:2008} provide metaheuristic quantum algorithms for such difficult combinatorial optimizations. 
 They utilize adiabatic evolution of quantum bits (qubits) to find the ground state of Ising spin-glass models. Since quantum-annealing processors (quantum annealers) have become available \cite{DWave:2011}, practical usage as well as fundamental researches on quantum optimization has largely been developed in recent years (see e.g. \cite{Albash:2018,Hauke:2019} and references therein).

Despite the great progress that has been taken place in the development of quantum optimization, the number of qubits 
as well as  the noise control are still limited.  To ulilize such 
noisy intermediate-scale quantum  (NISQ)  devices \cite{Preskill:2018}, hybrid systems
 that are capable of dealing with large-scale optimization problems while using relatively small quantum optimization need to be developed. 
So far, various hybrid algorithms have been proposed in the literature (see, e.g. \cite{ref-qbsolv,Chancellor,OKSolv,SamplePersistency_original, SamplePersistency,brief_propagation, QAGA, HybridCVRP, HybridDiscreteContin} and references therein). Most of them are based on the idea of decomposing original large-scale problem into subproblems to be treated by available quantum devices, so that multiple iterations between classical and quantum solvers are required,
while some are based on identification of a plausible subproblem by fixing persistent variables in multiple sampling of classical solvers \cite{SamplePersistency_original, SamplePersistency}.
 
\begin{figure}[h]
\begin{center}
\includegraphics[scale = 0.50]{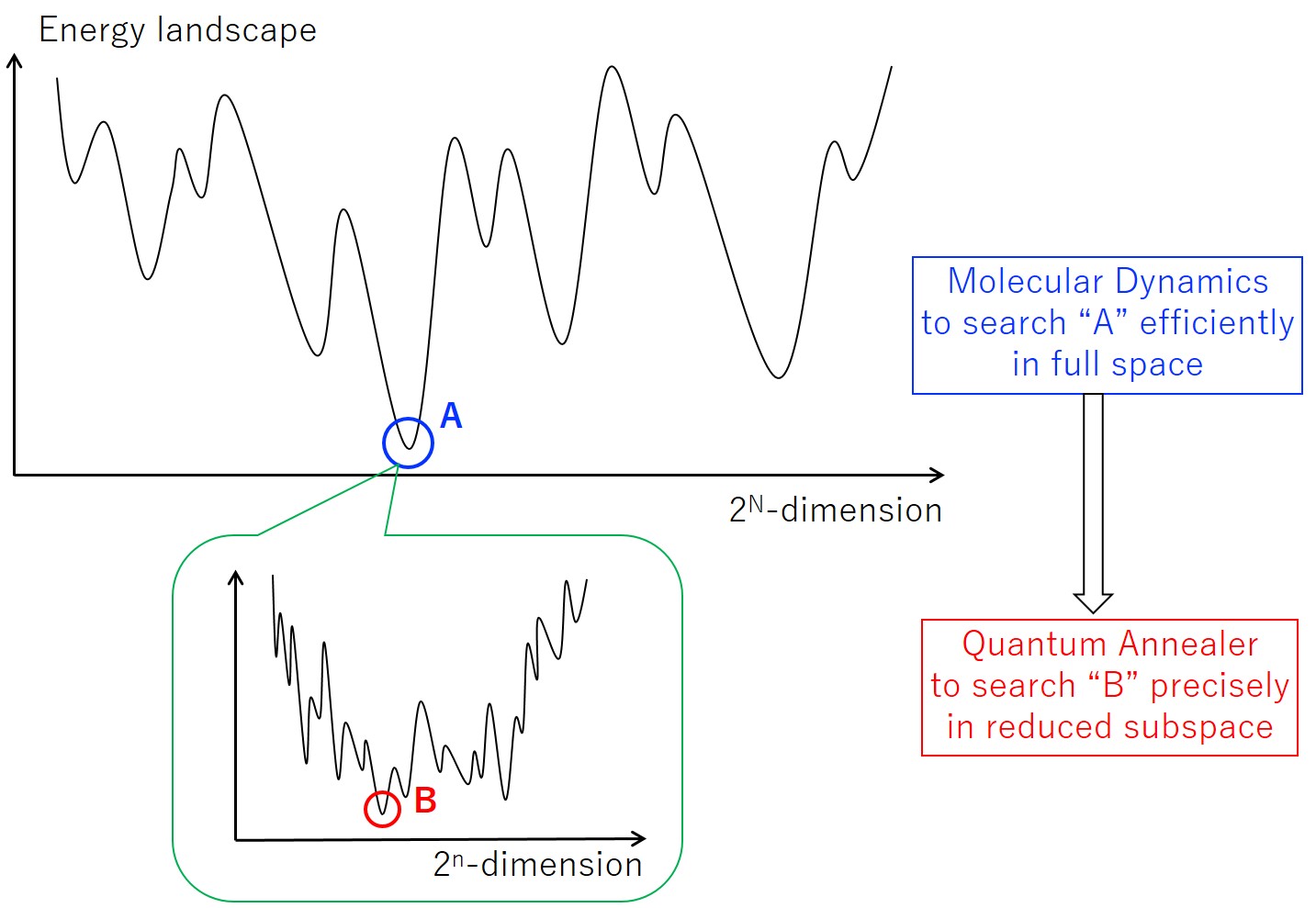}
\caption{\footnotesize Concept of  Hybrid Quantum Annealing  via Molecular Dynamics. \label{Fig-Concept}} 
\end{center}
\end{figure}

 In this paper,  we propose a novel hybrid system of quantum optimization, 
 Hybrid Quantum Annealing via Molecular Dynamics (HQA-MD, or HQA for short), 
 based on a combination of the classical molecular dynamics (MD)  \cite{CMD}  and the quantum annealing (QA) \cite{Kadowaki:1998}.
    The concept of HQA is illustrated in Fig.~\ref{Fig-Concept}. 
Consider the Ising spin-glass with $N$ number of sites.
 Only a single run of the  classical MD solver  with continuous flux variables  
is capable of  identifying a set of  low-energy spin configurations  with $2^n$-dimension indicated schematically by the region $A$ 
in  the full $2^N$-dimensional space, where $n$ is an arbitrarily chosen number smaller than $N$.
The quantum solver with quantum spin variables 
can then resolve the fine structure of 
the reduced $2^n$-dimensional subspace  to find the minimum $B$.
Thus, our classical MD solver acts as a powerful preconditioner 
to extract difficult 
spin variables automatically from the huge energy landscape and send them to the quantum annealer.

    For HQA to work in practice, it is crucial to develop suitable
     classical Hamiltonian dynamics. The molecular dynamics  of the classical fluxes can be used as a 
powerful preconditioner to sort out 
the ``frozen'' and ``ambivalent'' spins for quantum annealers, as we see below.
        Since both classical and quantum Hamiltonians have  the same roots,
        HQA constitutes a  seamless scheme for quantum-classical hybridization, 
so that some of the various developments that improve the performance of QA can also be imported into HQA and be utilized in large-scale Ising spin-glass models. 
   We note that the classical part of our HQA  has  some similarity  with 
    SVMC (Spin-vector Monte Carlo) \cite{Shin:2014},
    CIM (coherent Ising machine) \cite{CIM}, and      
    SBM (simulated bifurcation machine)  \cite{SBM}.  
Therefore, it is natural to expect that our 
hybrid scheme can be also
applied to such continuous optimizations schemes. 

\

\

\noindent
{\bf {\large Results}}\\

\noindent
 A large class of combinatorial optimization problems can be mapped onto  the 
  Ising model  
  \begin{align}
  {\cal H}_{\rm Ising} (s)= \frac{1}{2} \sum_{i\neq j}^N  J_{ij} s_i s_j + \sum_{i=1}^N h_i s_i,
  \label{eq:Ising}
  \end{align}
  with the Ising variables ($\{ s_i = \pm 1 \}_{i=1}^N$),  the 
symmetric  coupling ($J_{ij}$) and the external field ($h_i$) \cite{Lucas:2014}. 
 The quantum annealing (QA) of transverse-field Ising model \cite{Kadowaki:1998}
  provides an efficient method to solve the  ground state of the system through the
  quantum deformation of ${\cal H}_{\rm Ising}$ as  
\begin{align}
\mathcal H_{\rm QA}(\sigma;\tau)  =A(\tau)  \biggl[ - \sum_{i=1}^N \sigma_i^x \biggr]
+ B(\tau) \biggl[ \frac{1}{2} \sum_{i\neq j}^N J_{ij} \sigma_i^{z} \sigma_j^{z} 
+ \sum_{i=1}^N h_i \sigma_i^z \biggr], \label{eq_QA_Hamiltonian}
\end{align}
where  $\sigma_i^x, \sigma_i^z$ (and also $\sigma_i^y$) are 2$\times$2 Pauli matrices at each site $i \ (=1, 2, \cdots, N)$, 
 and $\tau$ is a fictitious time taken to be in an interval $[0, 1]$. 
 The scheduling functions
  $A(\tau)$ and $B(\tau)$ are chosen so that  ${\cal H}_{\rm QA}(\tau)$ interpolates   adiabatically
   the non-interacting spins with transverse field at initial time ($A(0) \gg B(0)$ )   and the classical Ising spin-glass  at final time ($A(1)\ll B(1)$).
     (The actual scheduling functions
    in our numerical experiments below are $A=A_{_{\rm DW}}/2$ and $B=B_{_{\rm DW}}/2$   where   
    $A_{_{\rm DW}}$ and $B_{_{\rm DW}}$ are the scheduling functions   given in Fig.2 of \cite{D-Wave-AB}.)
    In the actual quantum annealing devices, quantum Ising spin is realized by the superconducting flux qubits described 
   by a quantum Hamiltonian $\mathcal H_{\rm device}(\hat{\varphi}, \hat{p}  ; \tau)$  written by the flux operators $\hat{\varphi}_i$  and their 
    conjugates $\hat{p}_i$  with the  canonical commutation, $[\hat{\varphi}_j, \hat{p}_k ] = i \hbar \delta_{jk} $  (see e.g.~\cite{D-WaveQubits}).

\

\begin{figure}[t]
\begin{center}
\includegraphics[scale = 0.54]{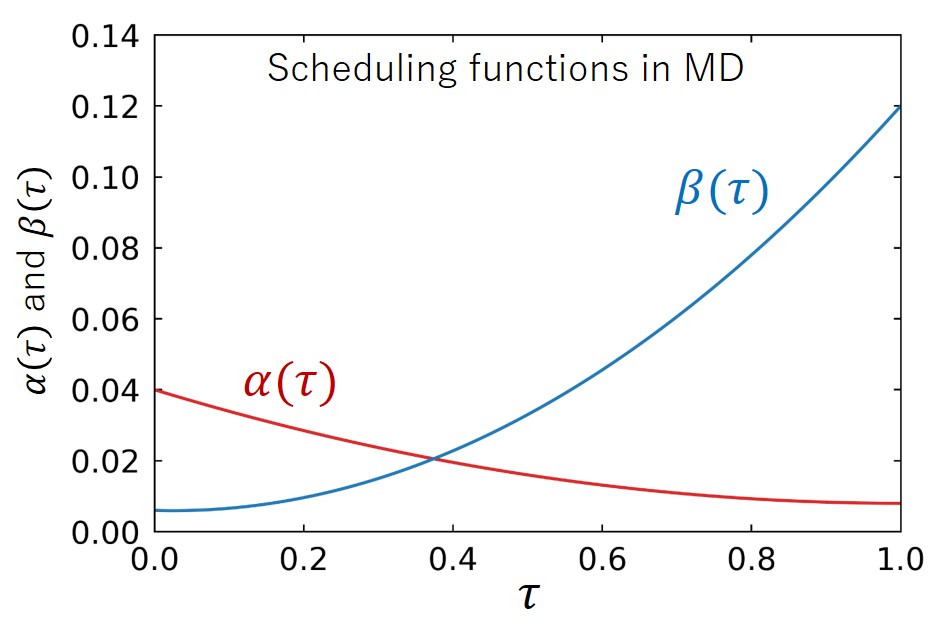}
\caption{\footnotesize The actual scheduling functions in  our MD run. See \blueflag{Methods} for their analytic forms.
\label{Fig-schedulingAB}} 
\end{center}
\end{figure}

\noindent
{\bf Molecular dynamics for flux variables}\\
To construct a seamless hybrid between 
quantum and classical solvers, 
 we introduce a classical Hamiltonian for flux variables:
 \begin{align}
\mathcal H_{\rm MD}(\varphi, p; \tau) = \alpha(\tau) \sum_{i=1}^N \Bigl( \frac{p_i^2}{2} + V(\varphi_i) \Bigr) 
+ \beta (\tau) \biggl[ \frac{1}{2} \sum_{i\neq j}^N J_{ij} \varphi_i \varphi_j
 + \sum_{i=1}^N 
 h_i |\varphi_i| \varphi_i \biggr] ,
 \label{Eq-MD-Hamiltonian}
\end{align}
where ``MD" stands for the Molecular Dynamics, $\{\varphi_i\}_{i=1}^N$  ($\{p_i\}_{i=1}^N$) 
are the continuous flux variables (continuous conjugate momenta)  which are classical counter parts of 
 $\{\hat{\varphi}_i\}_{i=1}^N$  ($\{ \hat{p}_i\}_{i=1}^N$).  
  The MD evolution is parametrized by $\tau=t/t_f \in   [0,1]$ with $t \in [0, t_f]$ being the actual evolution time.
    The potential term  $V(\varphi)$ is a convex downward function of the form $V(\varphi)= \varphi^M$ $(M=4, 6, 8, \cdots)$.
 Shown in Fig.~\ref{Fig-schedulingAB} are the actual scheduling functions ($\alpha(\tau)$ and $\beta(\tau)$) to be
  used in the present paper. The analytic forms are given in \blueflag{Methods}. 
 
 It is in order here to discuss the basic properties of the above classical Hamiltonian:
   The  term proportional to $\alpha(\tau)$ in Eq.(\ref{Eq-MD-Hamiltonian}) 
    ensures that each  classical flux variable oscillates 
    around $\varphi_i =0$ 
in early times.
  It  plays a similar role as the transverse-field term  proportional to $A(\tau)$ in  Eq.(\ref{eq_QA_Hamiltonian})
 which drives each spin state in early times to be  an equal superposition of up and down.
     The term proportional to $\beta(\tau)$  in  Eq.(\ref{Eq-MD-Hamiltonian}) 
    is a direct analogue of the Ising model: 
     By decomposing  the flux variables as
         $\varphi_i =  |\varphi_i|  {\rm sgn} ({\varphi}_i) $, one finds the 
      ``correspondence" between the terms  in
      Eq.(\ref{eq_QA_Hamiltonian}) and  Eq.(\ref{Eq-MD-Hamiltonian});
        $B J_{ij}  \leftrightarrow  \beta J_{ij}   |\varphi_i  \varphi_j |  $ 
      and   $ B h_{i} \leftrightarrow  \beta h_{i}   |\varphi_i \varphi_i|    $. 
     We note that   the classical dynamical system  achieves a faithful representation 
       of the Ising model, only when 
       all $|\varphi_i|$  are frozen to a positive constant $\mu$ and the equality $B=\beta \mu^2$ gets satisfied.
      However, this cannot be  achieved even for  ideal MD solvers,  and this is   a generic  problem  of all classical solvers using continuous dynamical systems.
    On the other hand, our MD solver plays a role of  a preconditioner for 
      the  quantum annealing, so that $\varphi_i$'s need not to settle down to $\pm \mu$.
   This is also the reason why  $\alpha(\tau=1)$ can be non-zero as shown in Fig.\ref{Fig-schedulingAB}.
 
 The Hamilton equations for the  time evolution of the flux variables reads
\begin{align}
 g \, \frac{d \varphi_i }{d\tau}= \frac{\del \mathcal H_{\rm MD}(\varphi, p;\tau)}{\del p_i},\qquad 
 g \,  \frac{d p_i  }{d\tau} = - \frac{\del \mathcal H_{\rm MD}(\varphi, p ;\tau)}{\del \varphi_i}, 
\label{eq:Hamilton-eq}
\end{align}
where $ \tau= t/t_f \equiv gt   $. The motion of the 
 flux variables becomes adiabatic for $g \rightarrow 0$. 
We solve the above equations by the  leapfrog algorithm ({\blueflag{Methods}}) on a  GPGPU machine.
 As the initial conditions,  we take  $\varphi_{i}(\tau=0)=0$, with
 $p_i(\tau=0) $ randomly chosen  to be   $+1$ or $-1$.
 As for the convex potential,  we have tested  $M=4, 6, 8$ and found that $M=6$ shows the best performance
  in terms of  the evolution time and the achieved accuracy, so that we use this value throughout this paper.

\begin{figure}[t]
\begin{center}
\includegraphics[scale = 0.60]{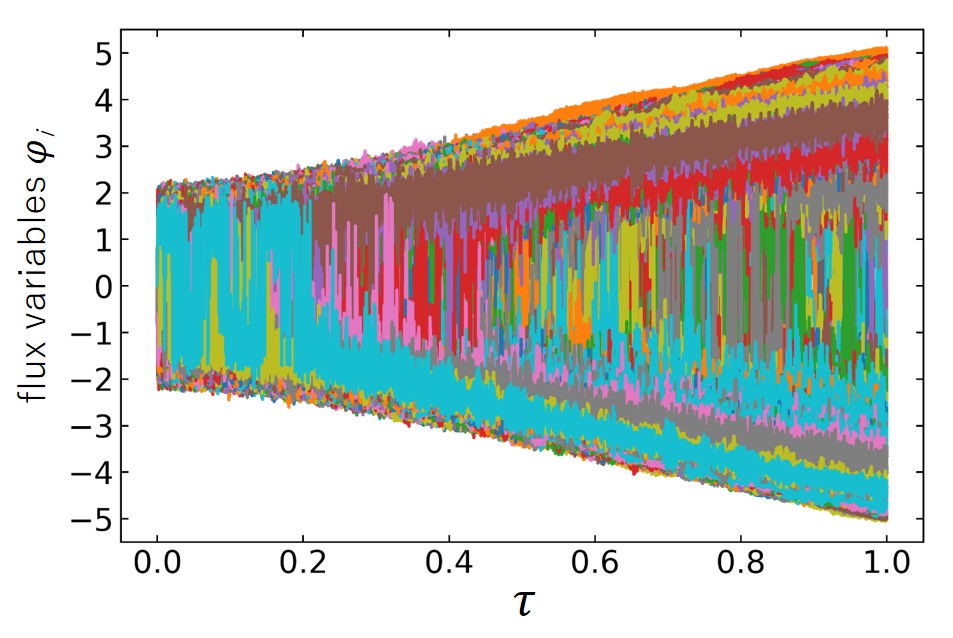}
\caption{\footnotesize Trajectories of all flux variables $\{\varphi_i \}_{i=1}^N$ 
 for a typical Ising spin-glass model
  with $N=10,000$  
and $(\delta \tau)^{-1}= 50,000$.
\label{Fig-full_traj}}
\end{center}
\end{figure}

\

\begin{figure}[t]
\begin{center}
\includegraphics[scale = 0.52]{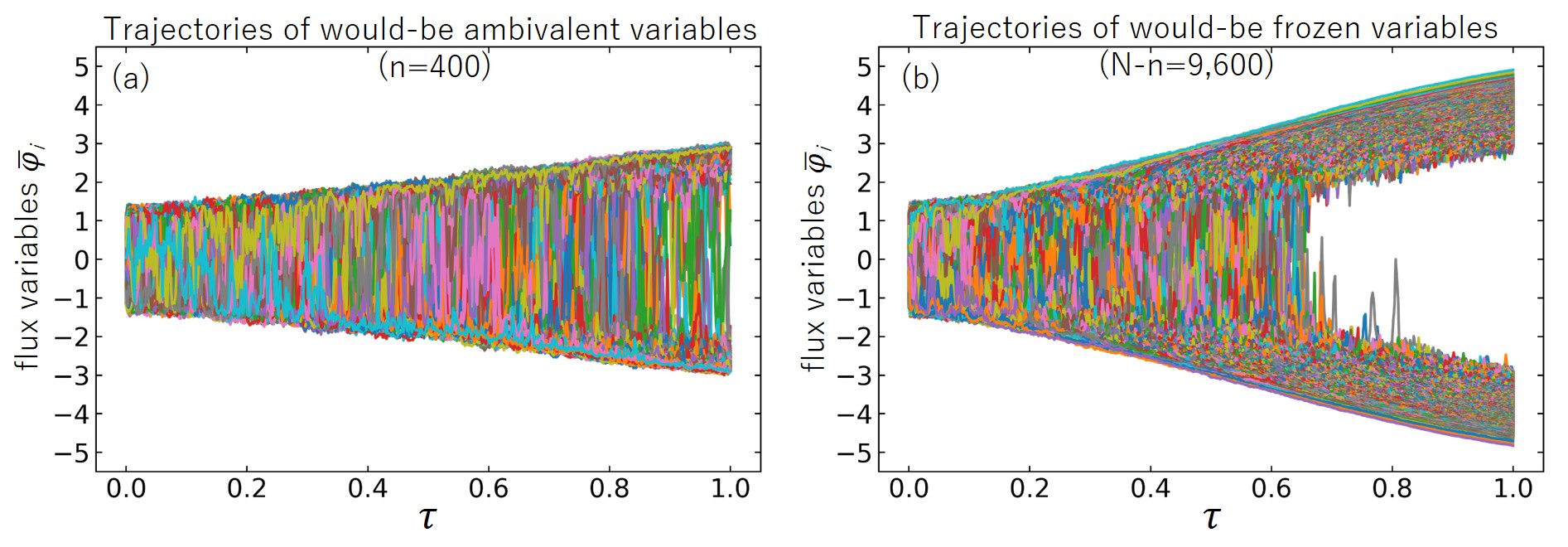}
\includegraphics[scale = 0.522]{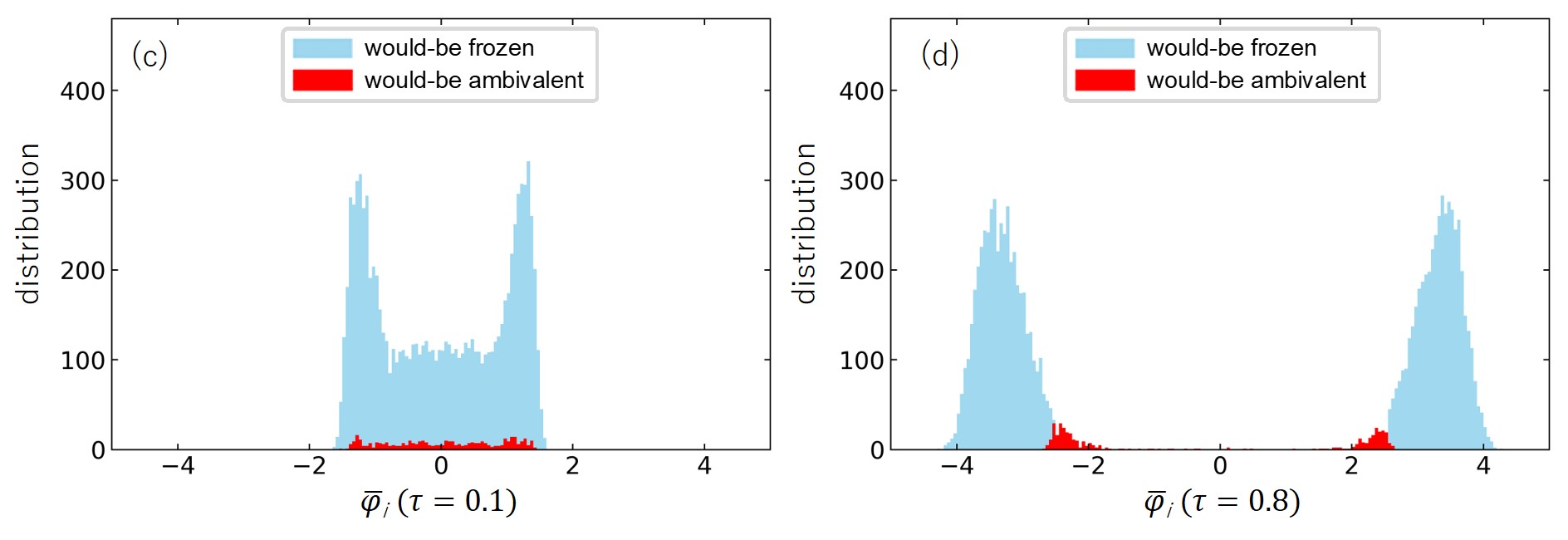}
\caption{\footnotesize (a) Trajectories of would-be ambivalent variables $\overline{\varphi}_{i'} $ with $i'=1,2,  \cdots, n$.
(b) Trajectories of would-be frozen variables  $\overline{\varphi}_{i'} $ with $i'=n+1,  \cdots, N$. 
Here, $n$ and $N$ are taken to be 400 and 10,000, respectively. 
Distributions of would-be frozen and ambivalent variables 
 at $\tau=0.1$ (c) and  at $\tau =0.8$ (d).
\label{Fig-Principle}}
\end{center}
\end{figure}

\noindent
{\bf  Sorting frozen and  ambivalent  variables}   \\
 Shown  in  Fig.\ref{Fig-full_traj} are all trajectories $\{ \varphi_i  \}_{i=1}^N$  ($N=10,000$) as a function of $\tau$  in   
      a test   MD evolution  with
a single set of Ising spin-glass parameters  picked up  randomly in  the intervals, 
   $-1 \le J_{ij} \le +1$    and   $-2 \le h_i \le +2$.  
 The MD time step  $\delta \tau$ is chosen to be 1/50,000. 
  Moreover, we make an identification, $g=\delta \tau$, so that the small time step corresponds to the adiabatic evolution.
   Although  there is  a  tendency that $\varphi_i$ fall into two categories with positive sign and negative sign, 
   we need a criterion to separate them in a quantitative manner.
   For that purpose, let us introduce 
 time-averaged flux variables,
\begin{align}
 \overline{\varphi}_i(\tau)  \equiv \frac{1}{\delta } \int^{\tau}_{\tau-\delta}  d\tau' \, \varphi_i(\tau'), 
\end{align}
  where the interval $\delta$ should be sufficiently larger  than $\delta \tau$ and sufficiently smaller than 1.
    Then all  trajectories can  be sorted by using their magnitudes at $\tau=1$ as 
$\bigl| \overline{\varphi}_{1'} (\tau=1) \bigr|  \le   \bigl| \overline{\varphi}_{2'} (\tau=1)
  \bigr| \le  \cdots   \le  \bigl| \overline{\varphi}_{n'} (\tau =1 ) \bigr| \le \cdots 
    \le  \bigl| \overline{\varphi}_{N'} (\tau =1 ) \bigr| $  where  $i'$ is  an index after the sorting and $n$ is an arbitrarily chosen integer 
    less than $N$.  Then, we call the $n$ low-lying trajectories,  $\left\{ \varphi_{i'}\right\}_{i'=1}^n$, as
     {\it ambivalent} variables,  and the rest is called  {\it frozen} variables.
 Shown in Fig.\ref{Fig-Principle}(a) with  $\delta = 100 \cdot \delta \tau$ are  the time-averaged
   trajectories $\left \{ \overline{\varphi}_{i'} (\tau) \right \} _{i'=1}^n$
 with $n=400$,  while Fig.\ref{Fig-Principle}(b)  shows  all the other 9,600 trajectories.  
  These figures indicate that 
   most of the flux variables are 
 frozen in sign  after the MD evolution, while 
     small number of  ambivalent variables remains  at $\tau=1$. 
  In Fig.\ref{Fig-Principle}(c) and (d), we show the distributions of the would-be  frozen and ambivalent variables
    at an early time ($\tau=0.1$) and at a late time ($\tau=0.8$).
    As  the time goes by, the distinction between two categories becomes prominent.

  \

\begin{figure}[t]
\begin{center}
\includegraphics[scale = 0.42]{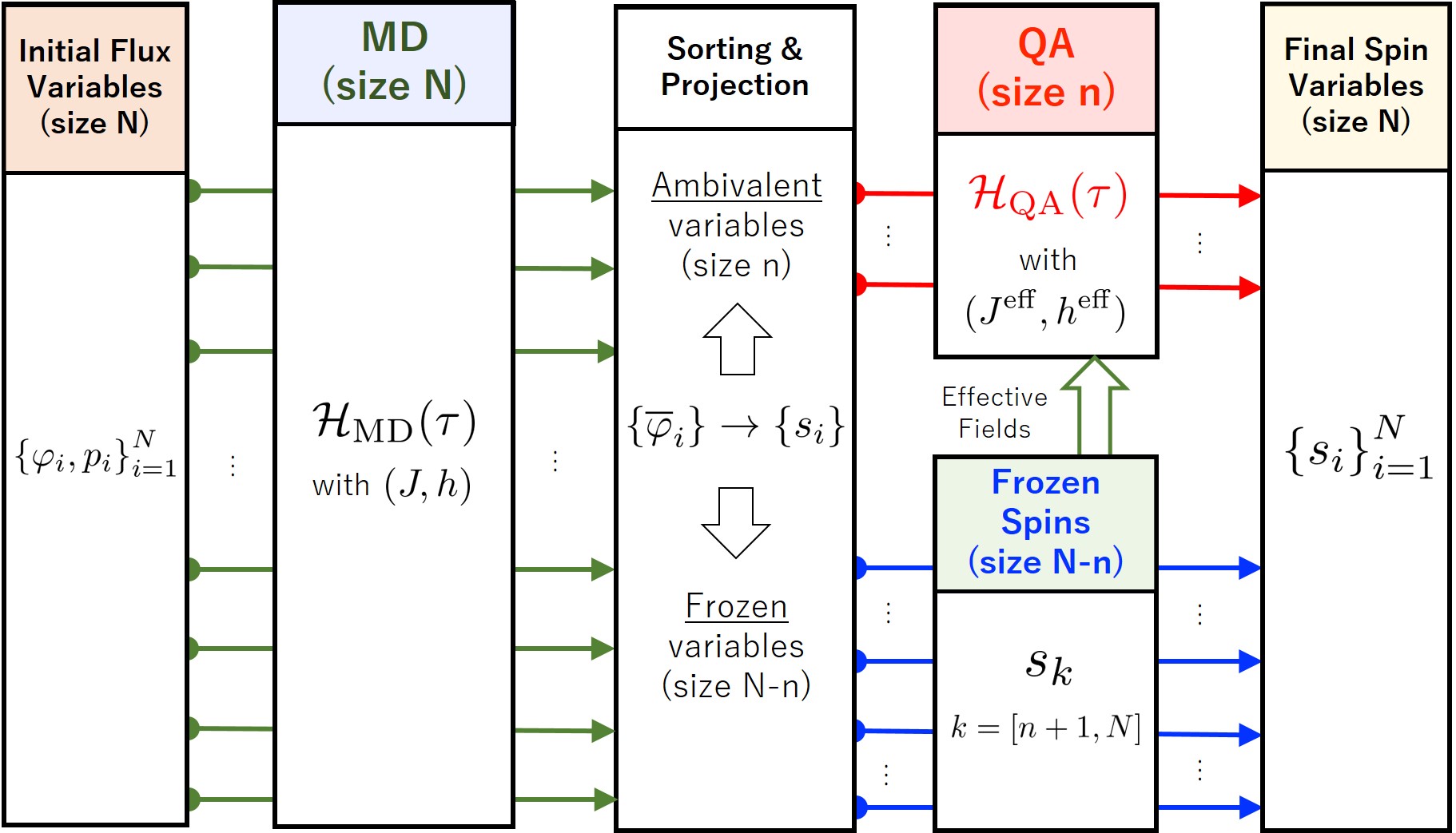}
\caption{\footnotesize Flowchart  of the Hybrid Quantum Annealing (HQA) via Molecular Dynamics (MD). \label{Fig-hybrid}} 
\end{center}
\end{figure}

\begin{figure}[tbp]
\begin{center}
\includegraphics[scale = 0.60]{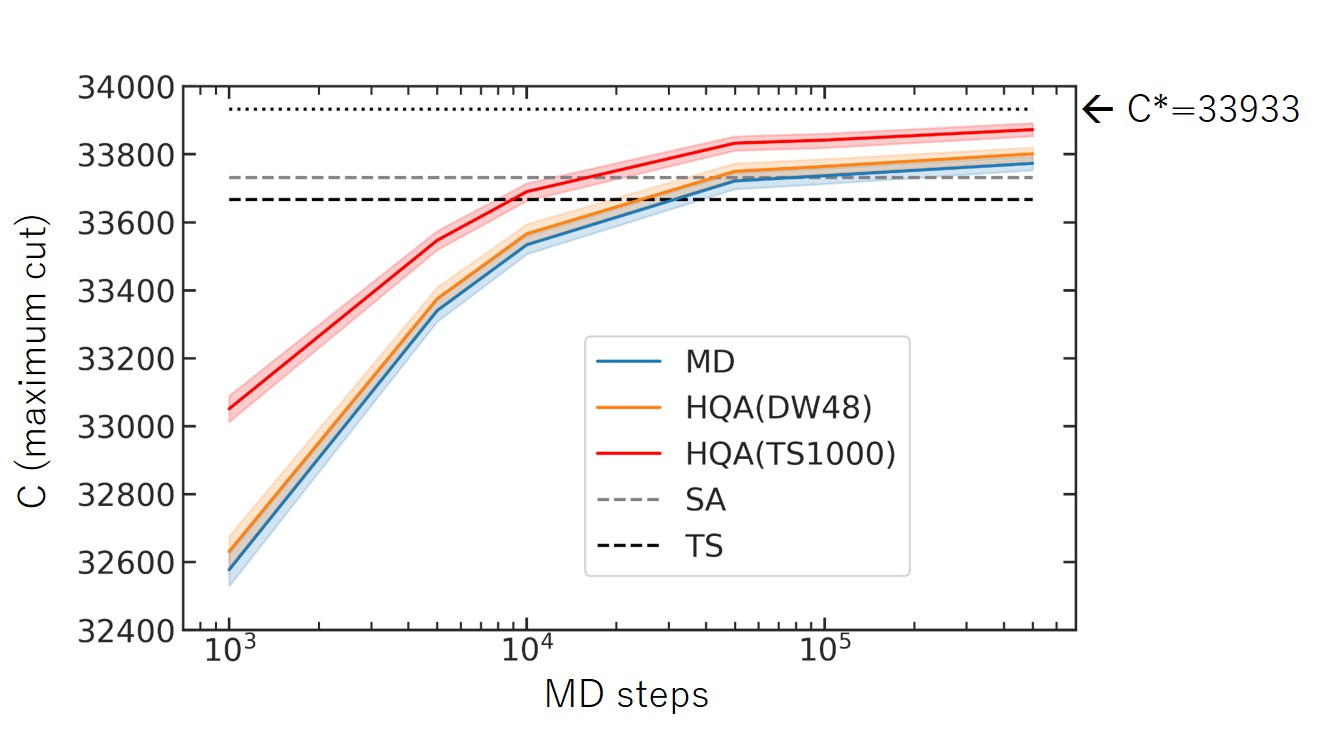}
\caption{\footnotesize The maximum cut $C$ in the MAX-CUT problem on a complete graph with 2000-node ($K_{2000}$)
 obtained by different solvers.   Theoretical estimate  of the  maximum cut is  $C^*=33,933$. \label{Fig-MaxCut}} 
\end{center}
\end{figure}

 \noindent
{\bf Hybrid Quantum Annealing via Molecular Dynamics}  \\
Our MD evolution combined with the above sorting algorithm can extract the 
 ambivalent variables efficiently.  
 For instance, as is shown in Fig.~\ref{Fig-MaxCut} (the line labeled by ``MD''), the obtained configurations successfully approach to the optimum. 
 However,   it takes  an exponential  time
   for those ambivalent variables to really  settle down, 
   and it is 
 not guaranteed to converge to the optimum. 
   Therefore, it is highly impractical 
  to continue the MD evolution toward  $\alpha =0$. Our approach to 
 circumvent this issue is  a novel hybrid scheme (HQA) where  MD is used as a powerful preconditioner for QA.
   Currently available  quantum annealers as well as quantum hybrid solvers are  still limited in  size and accuracy.
    Nevertheless,  as will be demonstrated below, the HQA
    complements the large-scale capability of such quantum solvers and enhances the performance in solving large $N$ optimization problems. 
     
     Our HQA is operated in the following way: 
We fix the frozen spins  ($k' =n+1, \cdots, N$)  by the projection
$s_{k'} = \sgn\bigl( \overline{\varphi}_{k'} (\tau =1) \bigr) $, while
 the ambivalent spins ($i' =1, \cdots, n$) are sent to a reduced size Ising subsystem 
  with the Hamiltonian,
\begin{align}
{\mathcal H'}_{\text{Ising}}(s) = \frac{1}{2}  \sum_{i' \neq j'} ^{n} J^{\rm eff} _{i'j'} s_{i'} s_{j'} +  \sum_{i'=1}^n h^{\rm eff}_{i'} s_{i'}  
 \equiv   {\mathcal H}_{\text{Ising}}(s|s_{k'=n+1, \cdots, N}:{\rm frozen} ) -  ({\rm const.}).
\end{align}
Here the effective couplings read
\begin{align}
 J^{\rm eff}_{i'j'} = J_{i'j'},  \ \ \ 
 h^{\rm eff} _{i'} = h_{i'} + \sum_{k'=n+1}^{N} J_{i' k'} s_{k'},  \ \ 
 \bigl(i', j' = 1, 2, \cdots, n \bigr) . 
\end{align}
This small subsystem of $n$ degrees of freedom can be solved by embedding it into a quantum annealer
or other Ising solvers. 
Shown in Fig.\ref{Fig-hybrid} is an overall  flowchart of our HQA  starting from initial flux variables
 $\{ \varphi_i, p_i \}_{i=1}^N$ 
and ending with  the final Ising-spin  variables  $\{ s_i \}_{i=1}^N$.

\

\noindent
{\bf HQA  for MAX-CUT problem}\\
 To demonstrate how our HQA works,
 let us consider the MAX-CUT problem which is to find the size of the maximum cut ($C$) in a given undirected graph.
 We take an all-to-all connected graph with 2000-node ($K_{2000}$)
 having the random bimodal edge-weight $w_{ij} = \pm 1 $ with zero-mean.
 This problem   has been  used for  benchmarking of various classical solvers including CIM \cite{CIM} and SBM \cite{SBM}.
 Mapping this problem into the Ising  model  (\blueflag{Methods}) with $J_{ij} = w_{ij}$, $h_i=0$ and $N=2000$,   
  we compare the performance of   three different cases; 
   our MD solver alone, HQA(DW48)  which is an  HQA with  the $n=48$ subsystem
  solved by the  D-Wave machine (DW\_2000Q\_5 \cite{D-Wave-AB}), 
      and HQA(TS1000) which is an HQA  with
        the $n=1000$ subsystem solved by the classical tabu search implemented as {\tt QBSolv} \cite{D-WaveQBSolv}.

For  reference classical solvers, we consider   the simulated annealing (SA)  ({\tt dwave.neal}  \cite{D-WaveSA}) and the  tabu search (TS). 
The number of sweeps of SA is set to be $N_{\rm sw}=$1000 with the inverse temperature $\beta$ increasing geometrically from  $\beta_{\rm I} = 0.01$ to $\beta_{\rm F} = 1.0$ at every single sweep. These values of $\beta_{\rm I, F}$ are chosen as a result of optimization over the random 100 instances. In our classical computational system (\blueflag{Methods}), the computational cost of $ N_{\rm sw} \times N (=1,000 \times 2000)$ SA steps  is comparable to that of the $10^6$ MD steps. TS in the present  study ({\tt QBSolv} implemented by D-Wave Systems, inc.) is already an optimized Tabu Search compared to a simple Tabu Search algorithm, and the computational time of TS is comparable to (or longer than) SA for more than a few thousands variables. 

In Fig.\ref{Fig-MaxCut}, the horizontal axis represents 
the number of computational steps $(\delta \tau)^{-1}$ in MD, while the vertical axis 
is the number of maximum cut ($C$) obtained by different solvers. 
Colored solid curves are the results of different solvers, MD, HQA(DW48) and HQA(TS1000). 
Note that the figure only shows dependence of MD steps. 
The band associated with each line represents $\pm 1\sigma$ confidence interval for 100 instances.  (In actual numerical experiments, each $J_{ij}$ is combined with a mirror instance $-J_{ij}$ to ensure $C_0 \equiv \frac{1}{4} \sum_{i \neq j} J_{ij} =0$.) 
In \blueflag{Supplementary Note 1}, the initial-condition dependence of MD and HQA  is shown with a single instance and 100 initial conditions. 
Theoretical value of $C$ using the finite size scaling analysis in statistical mechanics is
$C^* = -E^*/2  \simeq 33933$ \cite{SBM-eLetter} (\blueflag{Methods}) as shown by the dotted line. Here $E^*$ is the ground-state energy of the Ising model averaged over instances. 
The results of SA and TS with the above setting are shown by the gray dashed line and the black dashed line, respectively. 

From the figure,  one finds that the MD alone  reaches up to 0.4 $\%$ deviation from $C^*$  after $500,000$ MD steps.
This is more accurate  than the results of other classical solvers such as SA (0.6\% deviation) and TS (0.8 \% deviation) obtained under the comparable computational time.
Moreover, HQA shows further improvement of the solution toward $C^*$. Here with the same 500,000 MD steps, HQA(DW48) and HQA(TS1000) reach up to 0.3$\%$ and  0.2$\%$ accuracy, respectively. 
Note here that the primary computational time of HQA is consumed by the MD part in our computational systems (\blueflag{Methods}).
For example, the ratio of the computational time between DW48 and MD (500,000 MD steps) is 0.007, excluding the cloud connection to D-Wave machine.
Also, the ratio between TS1000 and MD (500,000 MD steps) is 0.24. 

If one continuously proceeds with the classical solvers (such as MD, TS or SA) to achieve the improvement the computational time will grow  substantially:  For example, to achieve 0.2 \% accuracy in SA, we find  it  necessary to increase the number of sweeps 10 times,  $N_{\rm sw}=10,000$.
Our HQA approach avoids such difficulty by extracting and solving  a computationally hard subproblem by quantum annealer or its quantum-classical hybrid systems. 
It is worth noting that even HQA(DW48) shows the improvement. For such a small system $(n=48)$, quantum annealer is not strong enough to be compared with   other classical solvers. However, it is still surprising that such a small subsystem improves the performance.

\

\begin{figure}[t]
\begin{center}
\includegraphics[scale = 0.49]{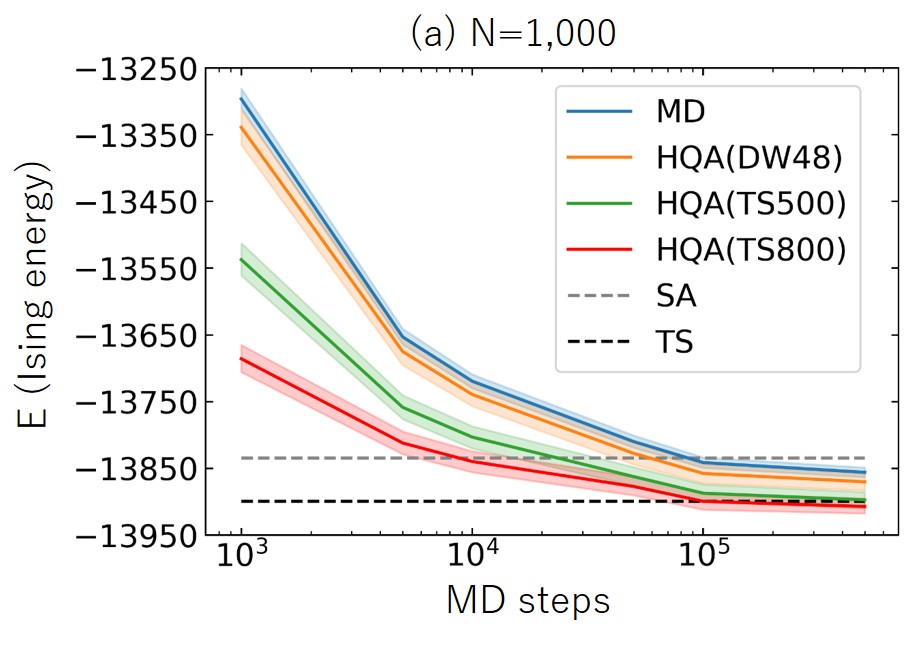}
\includegraphics[scale = 0.49]{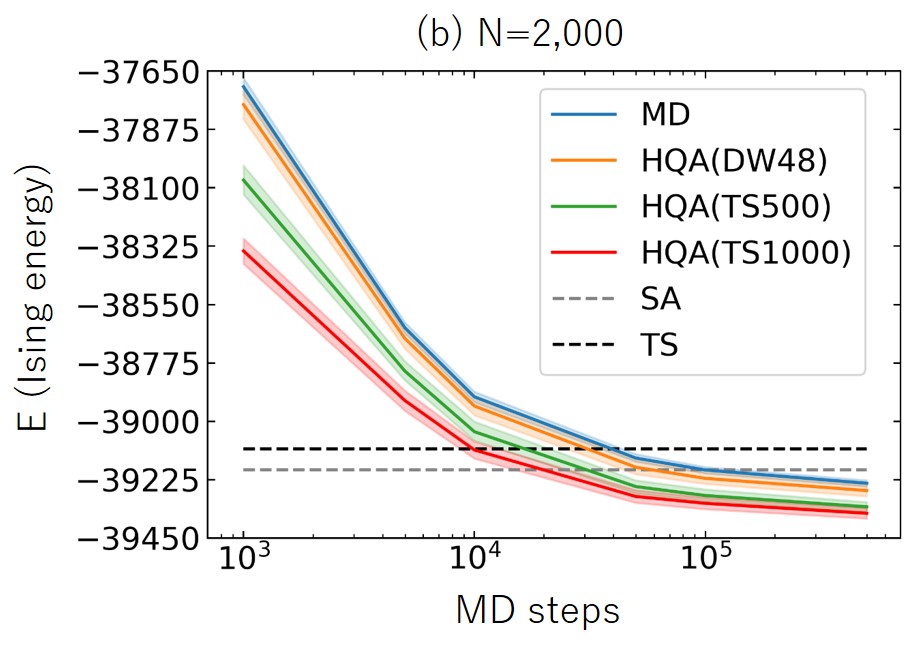}\\ 
\includegraphics[scale = 0.49]{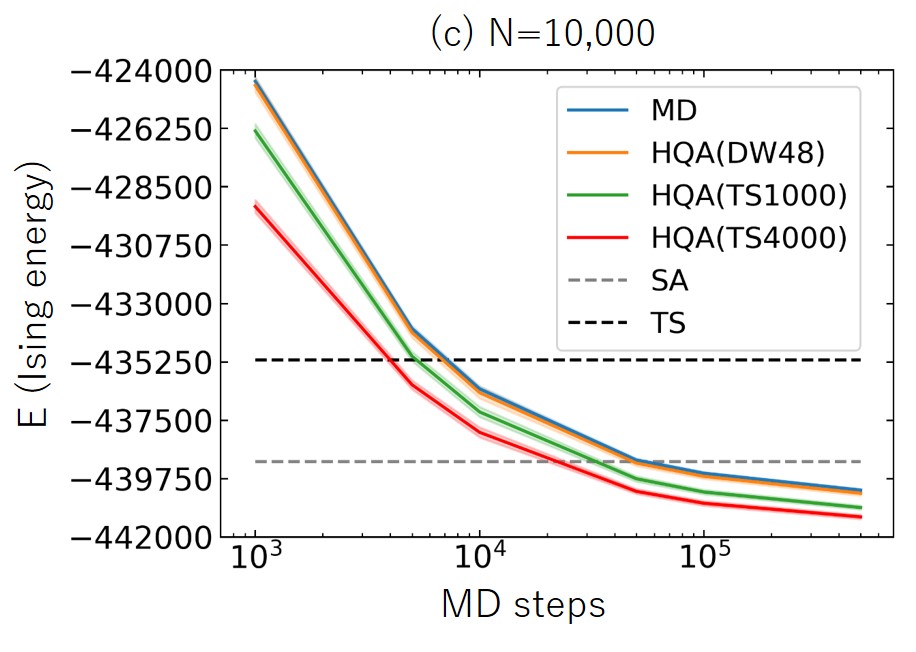} 
\caption{\footnotesize 
 Results of the Ising energy $E$ for the Ising spin-glass problem 
 averaged over 100 instances by using different solvers
  with three different system sizes  (a) $N=1,000$, (b) $N=2,000$, and (c) $N=10,000$.
\label{Fig-Benchmark}} 
\end{center}
\end{figure}

\noindent
{\bf HQA  for Ising spin-glass problem}\\
  Finally we consider a general Ising spin-glass model with 100 instances whose parameters $J_{ij}$ and $h_i$ 
  are randomly chosen in the interval  $-1 \le J_{ij} \le +1$ and   $-2 \le h_i \le +2$ (where the uniform distribution is utilized). 
  Total system sizes  are taken to be  $N=1,000$, $2,000$, and $10,000$ for several different values of  $n$
 in Fig.~\ref{Fig-Benchmark}(a,b,c).  
 Results of the  Ising energy averaged over instances  $E \equiv \vev{{\cal H}^{\rm (min)}_{\rm Ising}(s)}$
  are plotted as a function of the MD steps $(\delta \tau)^{-1}$ ranging from $1,000$ to $500,000$.
 (See \blueflag{Supplementary Note 2} for the adiabaticity of  MD evolution and its relation to the choice of the 
    scheduling functions.)
The colored solid curves are obtained by MD, HQA(DW48), HQA(TS500),  HQA(TS800),  HQA(TS1000)  and
  HQA(TS4000). 
   The  band associated with each line represents $\pm 1\sigma$ confidence interval for 100 instances.

The results of reference classical solvers, SA, and TS, are drawn by the gray and black dashed lines, respectively. 
The number of sweeps of SA is set to be $N_{\rm sw}=$1000 with the inverse temperature $\beta$ increasing geometrically from $\beta_{\rm I} = 0.01$ to $\beta_{\rm F} = 1.0$ at every single sweep. This value of $\beta$ is chosen as a result of optimization over random 100 instances for the Ising spin-glass model. 
Similar to the MAX-CUT cases,
 the computational cost of $ N_{\rm sw} \times N\, (=1,000 \times 1,000 \sim 1,000 \times 10,000)$ SA steps 
  is comparable to the $10^6\sim10^7$ MD steps. 
  Note that the computational time of TS is again comparable to (or longer than) SA
for more than a few thousands variables.
   
If the system size $N$  exceeds a few thousand, the accuracy of MD becomes  better 
than that of other classical solvers. 
For example, in the case of $N = 10,000$, the precision of SA with $N_{\rm sw}=1,000$  and $10,000$ can be obtained by MD alone with 50,000 and 500,000 MD steps, respectively.
Moreover, we find that HQA achieves better accuracy even further than the MD solution, where our MD solver acts as a powerful preconditioner to extract difficult spin variables even in the large-size problems.

\

\

\

\

\noindent
{\bf {\large Discussion}}\\

\noindent
In this paper, we introduced a quantum-classical hybrid scheme (HQA-MD, or HQA for short) which utilizes the molecular dynamics as a 
preconditioner for quantum annealing. By taking a classical Hamiltonian for flux variables associated with 
 spin variables, we have demonstrated that our HQA  can solve combinatorial optimization problems with  high accuracy.
    Moreover,  our HQA  shows better performance as  the system size becomes larger.
  There are various interesting  questions  to be studied further. Among others,
   generalization of HQA with non-stoquastic interactions needs to be developed  e.g. by adding  off-diagonal
   kinetic terms   in the MD solver, $\sum_{i<j} \ell_{ij} \, p_i p_j$ \cite{D-WaveNonStoq}.
   Moreover,  it is   important  to find  proper classical dynamics applicable not only to the $\mathbb Z_2$ spin variable
  but also to the  binary ($0$ and $1$)  and multi-valued variables. Also, 
   the algorithmic difference between our HQA (which preserves the adiabaticity from the beginning to the end)
   and SBM  \cite{SBM} (which breaks the adiabaticity at the point of bifurcation) should be clarified to understand 
  the role of  classical adiabaticity. It is also important to find a mathematical theorem which can quantify how close our MD solver can approach to the ground state. 
  With all these future works,  our quantum-classical hybrid scheme provides  a promising method to obtain efficient and precise solutions
  for  optimization problems in  science and technology.

\

\

\

\

\

\noindent
{\bf {\large Methods}}\\

\noindent
{\bf Computational systems}\\
For quantum annealing processor, we utilized the lower-noise D-Wave 2000Q quantum processor {\tt DW\_2000Q\_5} in our numerical experiments. The scheduling functions and the working graph of this processor is available in \cite{D-Wave-AB}. It enables us to embed the 48-node complete graph $K_{48}$ to this processor with the standard triangle clique embedding scheme (See e.g.~\cite{Boothby:2016}). Quantum annealing is conducted with {\tt chain\_strength = 15}, {\tt num\_reads = 10,000}, {\tt postprocess = `optimization'}, and {\tt annealing\_time = 20} [$\mu$sec].  The {\tt chain\_strength} parameter is optimized with random instances associated with the Ising spin-glass. 
For classical computation, we utilized a system composed of Intel Xeon Platinum 8260 CPU @ 2.40GHz (384GB memory) and NVIDIA TeslaV100 GPU (32GB memory). GPU acceleration is utilized in the case of MD calculations, for which parallel computation on GPU can be implemented in a straightforward manner.

\

\noindent
{\bf Scheduling functions for MD}\\ 
 We employ
$\alpha(\tau) = \alpha_f \bigl( \tau + \rho_1 (1- \tau) + \rho_2 \tau (\tau -1) \bigr)$
 and 
$\beta(\tau) = \beta_f \bigl( \tau +\kappa_1 (1- \tau)  + \kappa_2 \tau(\tau-1 ) \bigr)$,
with $(\alpha_f, \rho_1, \rho_2) = ( 0.008, 4, 3)$ and $(\beta_f , \kappa_1, \kappa_2)= (0.12,  0.05, 1)$. 
In early times when  $\alpha(\tau) \gg \beta(\tau)$, the flux variables $\{\varphi_i\}_{i=1}^N$ oscillate around $\varphi_i = 0$. This is a classical analogue of the initial quantum-superposition state of quantum annealing. 
If the motion of the flux variables  is sufficiently  faster than the evolution of scheduling functions, 
  the system approaches adiabatically to the final state  
  where 
   most of the flux variables $\{\varphi_i\}_{i=1}^N$ tend to be localized.

\

 \noindent
{\bf Leapfrog algorithm}\\ 
 The Hamilton equations in Eq.(\ref{eq:Hamilton-eq}) for $i=1, \cdots, N$ can be solved accurately by 
 the leapfrog algorithm    \cite{Thijssen}.
 With a given initial condition at $\tau=0$, $\{\varphi_i(0), p_i(0)\}$,
 we integrate the Hamilton equations
 with the step size $\delta \tau$ being identified with $g$  as follows:
 \begin{align}
   \begin{cases}
     p_i^{(m+ \frac{3}{2})} - p_i^{(m+\frac{1}{2})} &=
     \displaystyle
 - {\alpha}^{(m+1)} \left.  \frac{\del V (\varphi)} {\del \varphi_i} \right|^{(m+1)}
 - 2{\beta}^{(m+1)} \Bigl[ \frac{1}{2} \sum_{j=1}^N  J_{ij} \varphi_j^{(m+1)}  + h_i \bigl|\varphi_i^{(m+1)}\bigr|\Bigr] , \\[5mm]
 \varphi_i^{(m+2)} - \varphi_i^{(m+1)} & = {\alpha}^{(m+\frac{3}{2})} p_i^{(m+\frac{3}{2})},
\nonumber
 \end{cases}
 \end{align}
together  with the initial half step,
 $ p_i^{(\frac{1}{2})} = p_i^{(0)} 
      -  \frac{ \alpha^{(0)} }{2}  \left.  \frac{\del V (\varphi)} {\del \varphi_i} \right|^{(0)}
 - \beta^{(0)} \Bigl[ \frac{1}{2} \sum_{j=1}^N  J_{ij} \varphi_j^{(0)}  + h_i \bigl|\varphi_i^{(0)}\bigr|\Bigr] $ 
 and  $ \varphi_i^{(1)} = \varphi_i^{(0)}  + \alpha^{(\frac{1}{2})} p_i^{(\frac{1}{2})} $.
 Here $m$ denotes the temporal step with  $\tau=m \cdot \delta \tau$ ($m=0, 1, 2, \cdots $).
 Also, we introduced an abbreviated  notation,   $f_i^{(m)} \equiv f_i(m \cdot \delta \tau)$ and
 $f_i^{(m+\frac{1}{2})} \equiv f_i((m +\frac{1}{2})\cdot \delta \tau)$ with $f=\varphi$, $p$, $\alpha$ and $\beta$. 
     The leapfrog  integrator   has only $O((\delta \tau)^2)$ error and is essential for 
    our MD evolution to be accurate enough.
     (If the Hamiltonian does not have explicit $\tau$-dependence which is not the case in the present  situation, 
      this integrator has nicer properties such as the  time-reversibility and the symplectic property.)
 
 \

 \noindent
{\bf  MAX-CUT and Ising spin-glass} \\
 For a given undirected graph $\mathcal G=(\mathcal V, \mathcal E)$ with an edge-weight $\{w_{ij}\}_{(ij)\in \mathcal E}$, the MAX-CUT  is a problem of finding a partition of vertices, 
$\mathcal V = \mathcal V_+ \cup \mathcal V_-$  with  $\mathcal V_+ \cap \mathcal V_- = \emptyset$, which maximizes the sum of $w_{ij}$ connecting the two sets, $C\equiv \sum_{i \in \mathcal V_+, j \in\mathcal  V_-, (ij) \in \mathcal E} w_{ij}$. This
 can be mapped to the problem of  maximizing 
$C(s) = \frac{1}{2}\sum_{(ij) \in \mathcal E} w_{ij} {(1 - s_i s_j)}$ with 
respect to the Ising spin variables $s_i = \pm 1$.
One can rewrite $C(s)$ in terms of the Ising spin-glass model ($J_{ij}=w_{ij}$, $h_i=0$)  as 
 $C(s)= - \frac{1}{2}{{\cal H}_{\rm Ising}(s)} + C_0$, with 
 ${\cal H}_{\rm Ising}(s) = \frac{1}{2} \sum_{i\neq j}  J_{ij} s_i s_j$ and $C_0 \equiv \frac{1}{4}\sum_{i\neq j}  J_{ij}$.
Minimizing the Ising energy ${\cal H}_{\rm Ising}(s)$ corresponds to maximizing the cut configuration.
 The instances of our experiment are given on the $2000$-node complete graph $K_{2000}$ with randomly generated bimodal
  weights $J_{ij} = \pm 1$. 
 Therefore, the constant $C_0$ follows the normal distribution with zero-mean
  for large $N$. 
The ground-state energy averaged over instances, $E^* \equiv \vev{{\cal H}^{\rm (min)}_{\rm Ising}(s)}$ for
 the corresponding spin-glass model
has been discussed in \cite{SBM-eLetter}:
The  finite-size scaling implies 
$E^*/N^{\frac{3}{2}} \xrightarrow[N \to \infty ]{}   e_0 + A/ N^{\omega} $. 
Here $e_0 = -0.7631667265(6)$ is the Parisi energy \cite{Oppermann:2007},  while
 $\omega=2/3$ and $A = 0.70(1)$ are a conjectured value 
  and a fitted value, respectively, of the numerical data for finite $N$. 
 Combining all, the estimated value of the maximum-cut $C^*$ on $K_{2000}$ reads
$C^* \equiv -E^*/2 = 33933(4)$, which we  refer  in Fig.\ref{Fig-MaxCut}.

\

\ \

\noindent
{\bf \large Acknowledgments}\\
The authors would like to thank Takashi Tsuboi for valuable comments and discussions. H.I. would also like to thank Tadashi Kadowaki and Akira Miki for valuable comments and discussions. T.H. was partially supported by the grants, JST CREST JPMJCR1913
and JSPS KAKENHI  19K22032.  T.D. was partially supported by
``Priority Issue on Post-K computer" (Elucidation of the Fundamental Laws and Evolution of the Universe),
``Program for Promoting Researches on the Supercomputer Fugaku"
 (Simulation for basic science: from fundamental laws of particles to creation of nuclei)
and Joint Institute for Computational Fundamental Science (JICFuS).

\

\noindent
{\bf \large Author Contributions}\\
All authors contributed extensively to the work presented in this paper.

\newpage

\noindent
{\bf \large Supplementary Note} \ \\

\noindent
{\bf 1. Initial condition in HQA}  \\
In the main text, we studied the  MAX-CUT problem  on the $2000$-node complete graph $K_{2000}$, averaged over $100$ instances with a 
 single  initial condition, $\varphi_i(0)=0$ and $p_i(0)$ taken randomly from $\pm 1$.  
   To check the initial-condition dependence, 
   we consider MAX-CUT problem over a single instance with  100 initial conditions,   
   $\varphi_i(0) =0$ and   a set of $p_i(0)$ taken randomly from $\pm 1$. 

\begin{figure}[bh]
\begin{center}
\includegraphics[scale = 0.45]{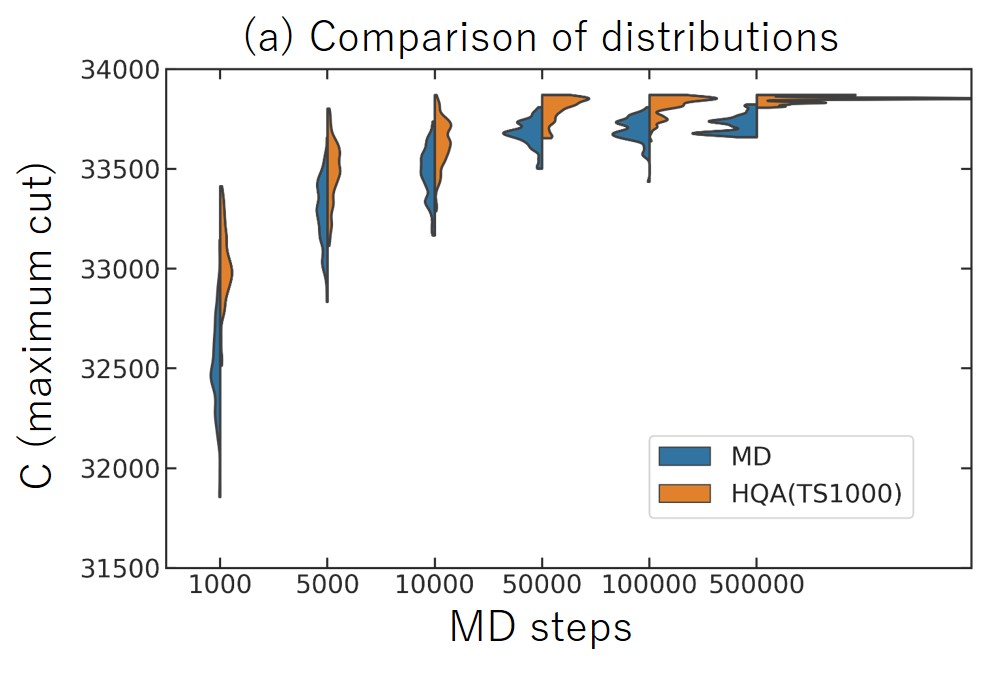}
\includegraphics[scale = 0.45]{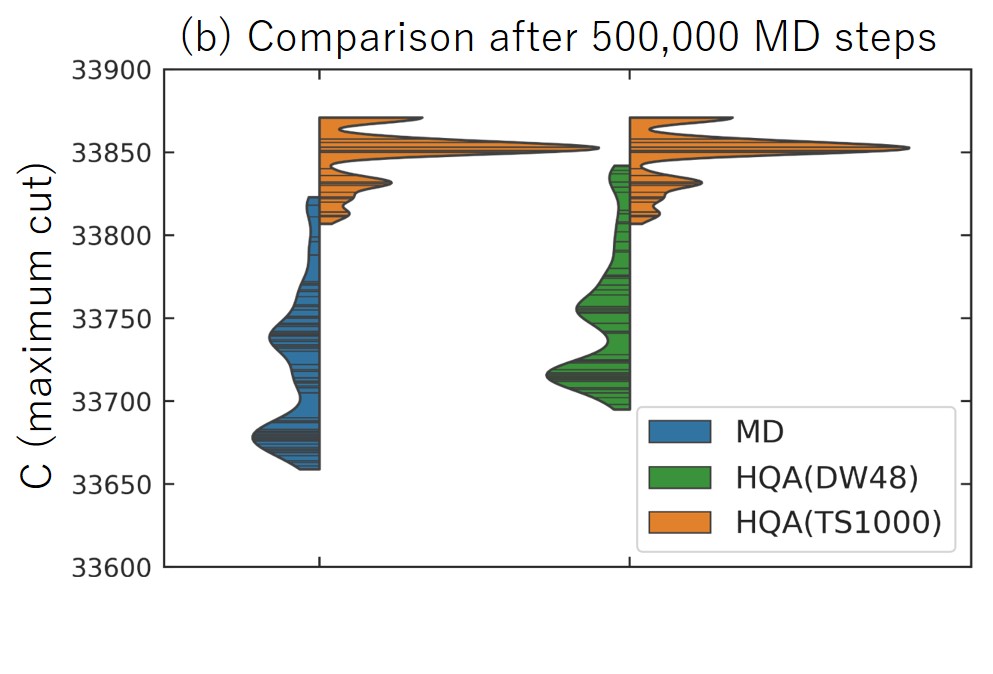}
\caption{\footnotesize (a) Distributions of $C$  associated with 100 initial conditions for the MAX-CUT problem on $K_{2000}$
  by using  the MD solver and  the HQA(TS1000) solver.
 (b) Comparison of the distribution of $C$  after $500,000$ MD steps among  MD, HQA(DW48) and HQA(TS1000). The horizontal lines in the distributions represent data points of 100 initial conditions.  \label{Fig-100-init}} 
\end{center}
\end{figure}

   Fig.~\ref{Fig-100-init}(a) shows the distribution of the maximum cut $C$ for a single instance with 100 initial conditions
       by using  the  MD solver and  the HQA(TS1000) solver.
   One finds that (i) the central value of the distribution increases as the MD steps increase (the same behavior as the 
    case of the 100 instances with a single initial condition), 
    (ii)   the distribution becomes sharper as the MD steps increase, and (iii) HQA provides larger $C$ with  
    sharper distribution than those of MD, which becomes prominent  for large MD steps.
  To  highlights the feature (iii) in a magnified scale, we show
  in    Fig.~\ref{Fig-100-init}(b) a comparison of the distributions of $C$ among MD, HQA(DW48) and  HQA(TS1000) after 
    500,000 MD steps.
\

\newpage

\begin{figure}[h]
\begin{center}
\includegraphics[scale = 0.48]{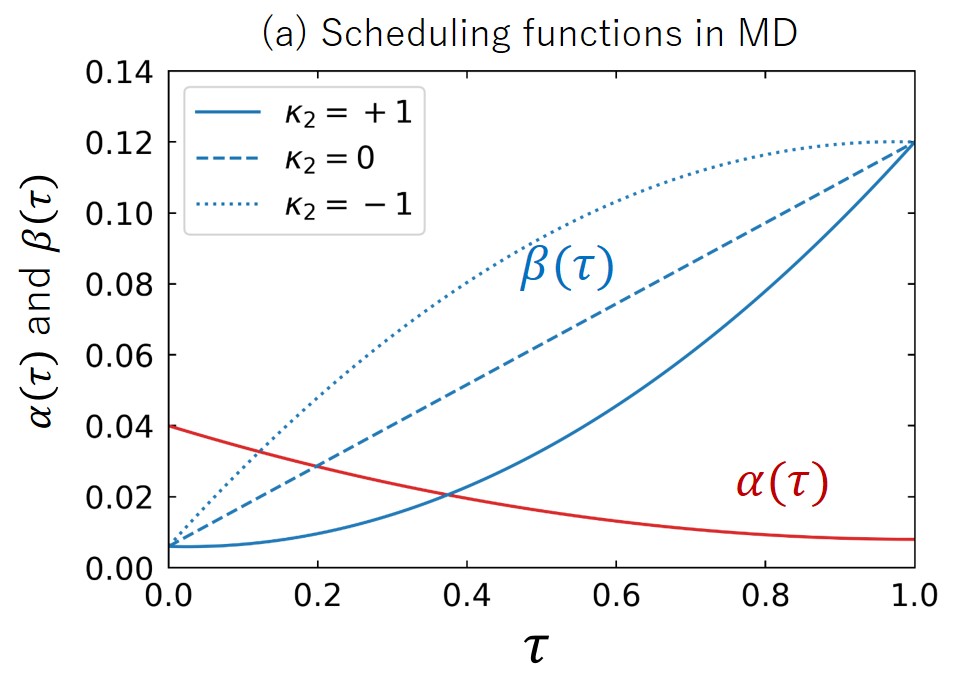}
\includegraphics[scale = 0.48]{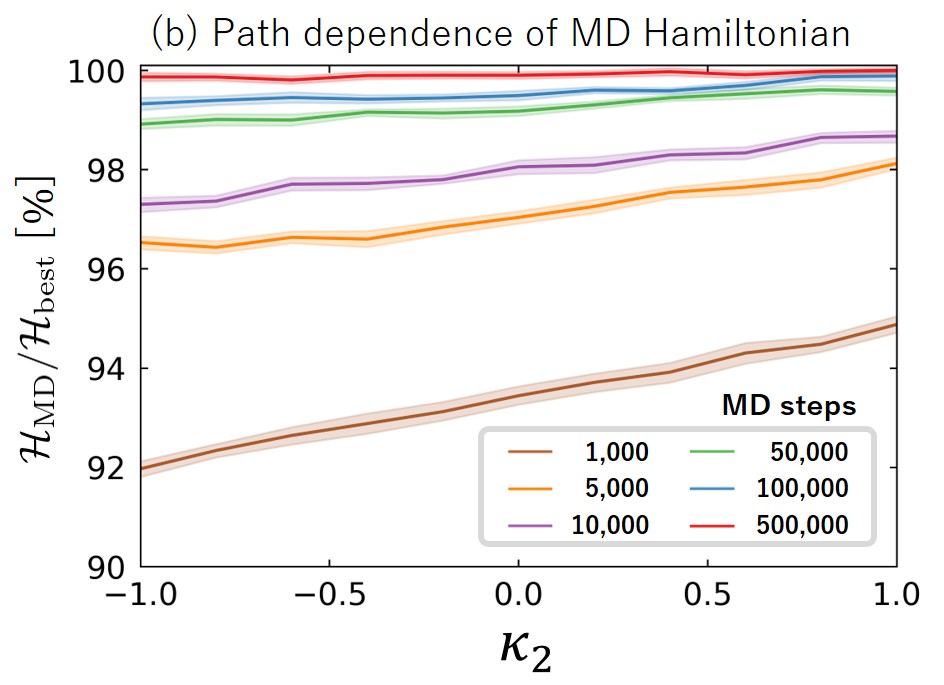}
\caption{\footnotesize (a) Typical three paths of thre scheduling function $\beta(\tau)$, parametrized in the interval  $-1 \leq \kappa_2 \leq 1$. (b) The 
 MD Hamiltonian at the end of the MD evolution ${\cal H}_{\rm MD}(\tau = 1)$ normalized by its best available value  ${\cal H}_{\rm best}(\tau = 1)$ 
 as a function of $\kappa_2$.  \label{Fig-Adiabaticity2}} 
\end{center}
\end{figure}

\noindent
{\bf 2. Adiabaticity in MD} \\
We have taken specific  forms of the scheduling functions $\alpha(\tau)$ and $\beta(\tau)$ for MD  in the main text.
 In this note, we study  the adiabaticity  of the MD evolution by changing one of the parameters $\kappa_2$ in 
$\beta(\tau) = \beta_f \Bigl( \tau + \kappa_1 (1- \tau) + \kappa_2  \tau (\tau-1) \Bigr)$.
 Keeping $\beta_f = 0.12$ and $\kappa_1 = 0.05$ as default values in the main text, we change 
  $\kappa_2$  within the interval  $[-1, 1]$ from the default value  $\kappa_2=+1$. We do not change 
  $\alpha(\tau)$ for simplicity. Shown in Fig.~\ref{Fig-Adiabaticity2}(a) are typical three paths of $\beta(\tau)$ as a 
  function of $\tau$.   
 
 With the above scheduling functions,
 we solve the Ising spin-glass problem of $N=10,000$ with a single instance and 10 initial conditions, and 
  calculate the final values of the  MD Hamiltonian at $\tau=1$ for  different  MD steps.
   In  Fig.~\ref{Fig-Adiabaticity2}(b), ${\cal H}_{\rm MD} (\tau=1)$ normalized by its best available value ${\cal H}_{\rm best}(\tau=1)$
    corresponding to  $\kappa_2=1$ and $(\delta \tau)^{-1}=500,000$ are plotted against $\kappa_2$.
 We find that (i) the value of 
 ${\cal H}_{\rm MD} (\tau = 1)$ is nearly independent of  $\kappa_2$ as long as the MD steps are  large enough, 
 indicating  that the adiabatic evolution is at work in our MD, and (ii) 
  the approach to ${\cal H}_{\rm best}(\tau = 1)$ is fastest for the default value $\kappa_2=+1$ employed in the main text.

\

\end{document}